\documentclass{article}

\usepackage[english]{babel}

\usepackage[letterpaper,top=2cm,bottom=2cm,left=3cm,right=3cm,marginparwidth=1.75cm]{geometry}
\usepackage[utf8]{inputenc}
\usepackage[T1]{fontenc}
\usepackage{multirow,multicol}
\usepackage{natbib}
\usepackage{amsmath}
\usepackage{graphicx}
\usepackage{subcaption}
\usepackage{mathtools}
\usepackage{amssymb}
\usepackage{amsthm}
\usepackage{bbm}
\usepackage{comment}
\usepackage{color}
\usepackage[colorlinks=true, allcolors=blue]{hyperref}
\usepackage[table]{xcolor}
\definecolor{lightgray}{gray}{0.9}
\usepackage{algorithm}
\usepackage{algorithmic}
\usepackage{amsmath}
\usepackage{graphicx}
\usepackage{subcaption} 
\usepackage{authblk} 
\newcommand{\performanceChangeSymbolNoX}{g}
\newcommand{\performanceChangeSymbolDifferential}{\;\text{d}\performanceChangeSymbolNoX}
\newcommand{\performanceChangeSymbol}[1]{g\left(#1\right)}
\newcommand{\performanceChangeSymbolMin}{g_{\min}}
\newcommand{\performanceChangeSymbolDistributionMean}[1]{\mu\left(#1\right)}
\newcommand{\performanceChangeSymbolDistributionVar}[1]{\sigma^2\left(#1\right)}
\newcommand{\performanceChangeSymbolDistributionStd}[1]{\sigma\left(#1\right)}
\newcommand{\performanceChangeSymbolDistributionShort}{\performanceChangeSymbolNoX}
\newcommand{\performanceChangeSymbolDistribution}{p\left(\performanceChangeSymbolNoX\,|\,\performanceChangeSymbolDistributionMean{\boldx}, \performanceChangeSymbolDistributionVar{\boldx}\right)}

\newcommand{\xdimsymbol}{m}
\newcommand{\boldx}{\textbf{x}}
\newcommand{\sample}{\textbf{s}}
\newcommand{\sampleRandVar}{\mathcal{S}}
\newcommand{\samplePdf}{p\left(\sample\right)}
\newcommand{\samplePdfLong}{p\left(\sampleRandVar=\sample\right)}
\newcommand{\population}{\mathcal{U}}

\newcommand{\datapoint}{\textbf{d}}
\newcommand{\differenceEstimator}{\hat{t}_{y,\text{dif}}}
\newcommand{\sampleDesign}{p(\sample)}
\newcommand{\Esampledesign}[1]{\mathbb{E}_{\sampleDesign}\left[#1\right]}
\newcommand{\EperformanceChange}[1]{\mathbb{E}_{\objectiveFunctionSymbol}\left[#1\right]}
\newcommand{\Indicator}[1]{\mathbbm{1}_{#1}}
\newcommand{\pseudocodeSymbol}[1]{\acquisitionFunctionSymb\left(#1\right)}
\newcommand{\pseudocodeSymbolMin}{\acquisitionFunctionSymb_{\min}}
\newcommand{\pseudocodeSymbolMax}{\acquisitionFunctionSymb_{\max}}
\newcommand{\horvitzThompsonEstimator}{\hat{t}_y}

\newcommand{\sumPopulation}{\sum_{\population}}
\newcommand{\sumSample}{\sum_{\sample}}

\newcommand{\Vsampledesign}[1]{\mathbb{V}_{\sampleDesign}\left[#1\right]}
\newcommand{\VperformanceChange}[1]{\mathbb{V}_{\performanceChangeSymbolDistributionShort}\left[#1\right]}

\newcommand{\standardSymbolMin}{z}
\newcommand{\standardSymbol}{u}
\newcommand{\standardDistributionShort}{\phi(\standardSymbol)}
\newcommand{\standardDifferential}{\;\text{d}\standardSymbol}
\newcommand{\standardDistributionFirstDer}{\phi'(\standardSymbol)}
\newcommand{\standardDistributionSecDer}{\phi''(\standardSymbol)}
\newcommand{\standardDistributionWithSymbol}[1]{\phi\left(#1\right)}

\newcommand{\ExpectationWRT}[2]{\mathbb{E}_{#1}\left[#2\right]}
\newcommand{\dataset}{\mathcal{D}}
\newcommand{\xpopulation}{\mathcal{X}}
\newcommand{\ypopulation}{\mathcal{Y}}
\newcommand{\datasetPrior}{\mathcal{D}_{\text{prior}}}
\newcommand{\datasetPosterior}{\dataset_{\text{posterior}}}
\newcommand{\xPrior}{\xpopulation_{\text{prior}}}
\newcommand{\yPrior}{\ypopulation_{\text{prior}}}
\newcommand{\NormalDistribution}[2]{\mathcal{N}\left(#1, #2\right)}
\DeclareMathOperator*{\argmax}{arg\,max}

\newcommand{\ExpectationGeneral}[2]{\mathbb{E}_{#1}\left[#2\right]}
\newcommand{\VarianceGeneral}[2]{\mathbb{V}_{#1}\left[#2\right]}
\newcommand{\ExpectationSymbol}[1]{\mathbb{E}\left[#1\right]}
\newcommand{\objectiveFunction}[1]{g\left(#1\right)}
\newcommand{\objectiveFunctionSymbol}{g}

\newcommand{\kernelfunction}[2]{k\left(#1, #2\right)}
\newcommand{\fmin}{\objectiveFunctionSymbol_{\min}}

\newcommand{\acquisitionFunctionSymb}{\alpha}
\newcommand{\acquisitionFunction}[1]{\acquisitionFunctionSymb\left(#1\right)}
\newcommand{\predictiveUncertainty}[1]{\text{PU}\left(#1\right)}
\newcommand{\lowerConfidenceBound}[1]{\text{LCB}\left(#1\right)}
\newcommand{\InvertedlowerConfidenceBound}[1]{\text{ILCB}\left(#1\right)}
\newcommand{\Improvement}[1]{\text{I}\left(#1\right)}
\newcommand{\ExpectedImprovement}[1]{\text{EI}\left(#1\right)}
\newcommand{\ScaledExpectedImprovement}[1]{\text{SEI}\left(#1\right)}

\newcommand{\x}{\textbf{x}}
\newcommand{\y}{\textbf{y}}
\newcommand{\w}{\textbf{w}}
\newcommand{\wpriorcov}{\Sigma_p}

\newcommand{\wpriormean}{\textbf{m}}
\newcommand{\s}{\sigma_{\varepsilon}^2}

\newcommand{\ytest}{\y_*}

\title{Model-assisted survey sampling with Bayesian optimization}
\author[1]{Jonne Pohjankukka}
\author[2]{Sakari Tuominen}
\author[1]{Jukka Heikkonen}

\affil[1]{Department of Computing, University of Turku, Vesilinnantie 5, 20500 Turku, Finland}
\affil[2]{Natural Resource Institute Finland, Latokartanonkaari 9, 00790 Helsinki, Finland}

\providecommand{\keywords}[1]{\textit{Keywords:} #1}

\begin{document}
\maketitle

\begin{abstract}
Survey sampling plays an important role in the efficient allocation and management of resources. The essence of survey sampling lies in acquiring a sample of data points from a population and subsequently using this sample to estimate the population parameters of the targeted response variable, such as environmental-related metrics or other pertinent factors. Practical limitations imposed on survey sampling necessitate prudent consideration of the number of samples attainable from the study areas, given the constraints of a fixed budget. To this end, researchers are compelled to employ sampling designs that optimize sample allocations to the best of their ability. Generally, probability sampling serves as the preferred method, ensuring an unbiased estimation of population parameters. Evaluating the efficiency of estimators involves assessing their variances and benchmarking them against alternative baseline approaches, such as simple random sampling. In this study, we propose a novel model-assisted unbiased probability sampling method that leverages Bayesian optimization for the determination of sampling designs. As a result, this approach can yield in estimators with more efficient variance outcomes compared to the conventional estimators such as the  Horvitz–Thompson. Furthermore, we test the proposed method in a simulation study using an empirical dataset covering plot-level tree volume from central Finland. The results demonstrate statistically significant improved performance for the proposed method when compared to the baseline.
\end{abstract}


\section{Introduction}

Survey sampling is a fundamental technique employed across various domains, such as environmental studies, social sciences, market research, medical research, agricultural studies and quality control to extract meaningful insights from a larger population of data.  It is a common tool used e.g. in natural resource monitoring which encompass ecosystems, wildlife populations, water bodies, mineral deposits, and more \citep[e.g.,][]{Stahl2016,Rabe2022,Danz2005,Abzalov2016}. Survey sampling plays a pivotal role in obtaining accurate and unbiased information about these resources, enabling informed decision-making for conservation, sustainable management, and policy formulation. In an era where data is hailed as the lifeblood of informed decision-making, the importance of sound and effective sampling techniques cannot be overstated. The fundamental premise of sampling lies in its ability to provide a cost-effective means of collecting a finite number of sample data that represents the larger population, potentially consisting from infinite number of sample points, as accurately as possible. The primary performance criteria for survey sampling methods typically revolve around their capacity to generate unbiased estimates of population parameters with the sample data, such as the population mean, while also optimizing efficiency, typically quantified by the variance of the corresponding estimator \citep{model2003sarndal}. In essence, the estimator's variance gauges how consistently close the estimator is to the true population value when, on average, it accurately estimates the population. Minimizing the variance of these estimators is often a primary objective, aiming to achieve what is known as the minimum-variance unbiased estimator \citep[MVUE, see e.g.,][Ch. 7]{Hoel2009} whenever feasible for the task at hand. 

Often times, survey sampling is combined together with wall-to-wall remote sensing (RS) data where a target population variable of interest (e.g. soil attributes) is estimated over a wider range using the sampled data and some statistical model \citep[e.g.,][]{STEHMAN1996169,Raty2018}. RS has emerged as a crucial tool in survey sampling by leveraging comprehensive data obtained from various sources inaccessible through on-ground sampling methods. By utilizing RS data, surveyors can extrapolate and infer trends, characteristics, and variables over large and remote areas, augmenting the limitations of on-ground sample collection. This approach enables a more comprehensive understanding of landscapes, ecosystems, and environmental changes, thereby enhancing the accuracy and scope of survey sampling applications in areas where obtaining sample data from every location is unfeasible. The evolution of RS and geographic information systems (GIS) in environmental monitoring has shown steady growth in research interest and their varied applications across fields such as water, air, soil, agriculture, and urban monitoring \citep{Lorena2022}. Additionally, the utilization of auxiliary RS data can play a useful role in augmenting the effectiveness of survey sampling estimators. By thoughtfully incorporating auxiliary data, survey sampling techniques are empowered to extract more precise and comprehensive insights, ultimately elevating the overall efficiency of the estimation process. \citep[e.g.,][]{LPM2011,MCROBERTS2022113168}

Sampling methods in statistical research can be broadly categorized into three main approaches: design-based, model-assisted, and model-based sampling. Design-based sampling entails selecting a probability sample from the response population using an appropriate sampling design. This sample is then utilized to create an unbiased estimator for a population parameter of interest \citep{Stahl2016}. Model-assisted sampling integrates design-based principles with statistical models, incorporating auxiliary information such as covariates to enhance estimator precision. This approach aims to improve sampling design efficiency while ensuring random and unbiased sample selection \citep{model2003sarndal}. In both design-based and model-assisted frameworks, the response population elements $y$ are considered unknown constants, and randomness arises from the sampling procedure itself. Conversely, in model-based sampling, the population elements $y$ are treated as random variables following a statistical model, for instance, a standard linear regression model with Gaussian noise and auxiliary variables such as wall-to-wall known RS data as model covariates \citep{Raymond2013}.

A methodology from optimization context with similar goals as survey sampling arises from the Bayesian framework called Bayesian optimization (BO). BO is a probabilistic approach to global optimization that aims to find the optimal solution of an objective function with minimal computational resources by iteratively evaluating the function at informative points. It combines probabilistic models, such as Gaussian processes, with acquisition functions to balance exploration and exploitation in the search space, efficiently navigating towards the optimal solution \citep{garnett_2023}. BO has become valuable method when making efficient decisions for subsequent optimization steps, particularly evident in tasks like model calibration. This is notably relevant in neural network context \citep[e.g.][]{NIPS2016}, where tailoring the network architecture to a specific application is crucial. Rather than resorting to inefficient random search or an exhaustive grid search of vast hyperparameter combinations, an incredibly resource-intensive approach, BO provides a solution by leveraging prior knowledge about successful steps in the past, allowing for sequential updates of strategies for subsequent steps \citep{WU201926}.

From the preceding discussion, we can distill the primary objectives of survey sampling and Bayesian Optimization (BO) into two main goals: 1) The development of unbiased sample-based estimators with minimal variance, and 2) The execution of effective sampling decisions to optimize a metric of interest. Given the two frameworks and shared objectives of survey sampling and BO, it prompts a question: Can we incorporate BO into a survey sampling framework by producing an optimized unbiased sampling design? This question forms the crux of our study, and we aim to address this issue by integrating the unbiased difference estimator from the model-assisted sampling framework with BO. From a BO perspective, our goal is to minimize the variance of population parameter estimators within a model-assisted sampling framework. A previous study by \citep{Pohjankukka2022mdpi} conducted a similar investigation, but it lacked a connection between the theoretical frameworks of survey sampling and BO. In this work, we strive to establish a more robust link.

\section{Methodological framework}
In this section, we will describe the methodological components which function as the basis of our novel method. We refer to the independent explanatory variables (known also as auxiliary data) as a $\xdimsymbol$-dimensional vector $\boldx\in\xpopulation\subset\mathbb{R}^\xdimsymbol$ and the corresponding target or response variables as $y\in \ypopulation \subset \mathbb{R}$. A single observed data point, with index $k$, is denoted as the tuple $\datapoint_k = (\boldx_k, y_k)$ and a sample of data points of size $n$  as $\sample = \{\datapoint_i\}_{i=1}^n \subset \population$, where $\population =(\xpopulation, \ypopulation)$ is the data population, that is, the set of all possible data points. Furthermore, in this paper we assume, which in practice often is the case, that the set of explanatory variables $\xpopulation$ (e.g. RS data) are easily available to us, but $\ypopulation$ (e.g. geographical soil samples) is not. The sample $\sample$ is a realization of a random variable $\sampleRandVar$ which has probability distribution $\samplePdf=\samplePdfLong$, called a \textit{sampling design} \citep{model2003sarndal}, over all possible samples. By $I_k$, we denote the binary random variable indicating that data point $\datapoint_k$ is in the sample $\sample$. In other words, $I_k=1$ or $I_k=0$ if $\datapoint_k\in\sample$ or $\datapoint_k\notin \sample$ respectively. By symbol $\pi_k =\Esampledesign{I_k}$ we denote the probability that data point $\datapoint_k$ is included in the sample $\sample$, and by $\pi_{kj}=\Esampledesign{I_kI_j}$ we denote inclusion probability that both data points $\datapoint_k$ and $\datapoint_j$ are included in the sample $\sample$. The sum notations $\sumPopulation$ and $\sumSample$ denote sum over all data points in the population and sample respectively.

\subsection{Gaussian process regression}
As a surrogate model in this work, we use Gaussian processes \citep[GP,][]{Rasmussen2005} for modeling real-valued functions as random GPs $\{f(\boldx), \boldx\in\mathbb{R}^\xdimsymbol\}$. At its core, a GP defines a distribution over functions, where any finite set of function values has a joint Gaussian distribution. Due to this property, GPs offer not only the capability to make point predictions but also model the uncertainties in those predictions. In GP regression (GPR), a set of input-output pairs (training data) is used to infer the underlying function's behavior. The GP defines a prior distribution over functions, and as new data points are observed, the prior is updated to a posterior distribution that reflects the learned function and its uncertainty. In other words, in a GP each point $(\boldx, f(\boldx))$ has a Gaussian distribution $f(\boldx) \sim \mathcal{N}(\mu(\boldx), \sigma^2(\boldx))$ in which the mean and variance components are functions of the corresponding explanatory variables $\boldx$. The observed response variable is modeled as $y = f(\boldx) + \varepsilon$ where $f(\textbf{x})=\phi(\boldx)^T \textbf{w}$ and $\phi(\boldx): \mathbb{R}^{\xdimsymbol} \to \mathbb{R}^p$. We assume the model weights $\w\in\mathbb{R}^{p\times 1}$ and noise term $\varepsilon\in\mathbb{R}$ are normally distributed, that is $\textbf{w}\sim \mathcal{N}(\wpriormean, \wpriorcov)$ and $\varepsilon \sim \mathcal{N}(0, \sigma_{\varepsilon}^2)$ where $\wpriormean\in \mathbb{R}^{p\times 1}$, $\wpriorcov\in \mathbb{R}^{p\times p}$ and $\sigma_{\varepsilon}^2>0$. Next, denote an observed data set of size $n$ as $\mathcal{D}=\{(\boldx_1, y_1), (\boldx_2, y_2), ..., (\boldx_n, y_n)\}$, $X=\{\boldx_1, \boldx_2, ..., \boldx_n\}, \y=[y_1, y_2, ..., y_n]^T$ and a set of $q$ observed explanatory variable points as $X_*=\{\boldx_1^*, \boldx_2^*, ..., \boldx_q^*\}$. If we now define the matrices $\Phi, \Phi_*$ in the following way: 

$$\Phi = \Phi(X) = \begin{bmatrix}\phi(\x_1), \phi(\x_2), ..., \phi(\x_n)\end{bmatrix} = \begin{bmatrix}\phi_{11} & \phi_{12} & \cdots & \phi_{1q} \\ \phi_{21} & \phi_{22} & \cdots & \phi_{2q} \\ \vdots & \vdots & \vdots & \vdots \\ \phi_{p1} & \phi_{p2} & \cdots & \phi_{pn}\end{bmatrix} \in \mathbb{R}^{p\times n}$$

$$\Phi_* = \Phi(X_*) = \begin{bmatrix}\phi(\x_1^*), \phi(\x_2^*), ..., \phi(\x_q^*)\end{bmatrix} = \begin{bmatrix}\phi_{11}^* & \phi_{12}^* & \cdots & \phi_{1q}^* \\ \phi_{21}^* & \phi_{22}^* & \cdots & \phi_{2q}^* \\ \vdots & \vdots & \vdots & \vdots \\ \phi_{p1}^* & \phi_{p2}^* & \cdots & \phi_{pq}^*\end{bmatrix} \in \mathbb{R}^{p\times q},$$

then the joint Gaussian posterior predictive distribution of $\y^* = \{y_1^*, y_2^*, ..., y_q^*\}$ given $X_*$ and $\mathcal{D}$, that is $p(\y^* | X_*, \mathcal{D})$, has an expected value $\mu_{y^*}$ and covariance $\Sigma_{y^*}$ given by:
\begin{equation}\label{Equation::posterior_predictive_mean_long}
    \mu_{y^*}=\mathbb{E}\left[\ytest|X_*, \mathcal{D}\right] = \Phi_*^T\wpriormean + \Phi_*^T\wpriorcov\Phi\left[\Phi^T\wpriorcov\Phi + \s I\right]^{-1}\left(\y-\Phi^T\wpriormean\right)
\end{equation}
\begin{equation}\label{Equation::posterior_predictive_covariance_long}   
\Sigma_{y^*}=\text{Cov}\left[\ytest|X_*,\y,X\right] =\s I + \Phi_*^T\wpriorcov\Phi_* -\Phi_*^T\wpriorcov\Phi\left[\Phi^T\wpriorcov\Phi +\s I\right]^{-1}\Phi^T\wpriorcov\Phi_*.
\end{equation}

It is common to denote $K(X, X)=\Phi^T\wpriorcov\Phi, K(X_*, X)=\Phi_*^T\wpriorcov\Phi, K(X, X_*)=\Phi^T\wpriorcov\Phi_*$, $K(X_*, X_*)=\Phi_*^T\wpriorcov\Phi_*$ and assume a prior zero mean $\wpriormean=\textbf{0}$ for the model weights $\w$, in which case the mean and covariance can be written as:

\begin{equation}
    \mu_{y^*} = K(X_*, X)\left[K(X, X) + \s I\right]^{-1}\y
\end{equation}
\begin{equation}    \Sigma_{y^*}=\s I + K(X_*, X_*) -K(X_*, X)\left[K(X, X) +\s I\right]^{-1}K(X, X_*).
\end{equation}

Notice that when $\w \sim \mathcal{N}(\textbf{0}, \wpriorcov)$ we have $\ExpectationWRT{\w}{f(\boldx)} = \phi(\boldx)^T\ExpectationWRT{\w}{\w}=0$ and $\ExpectationWRT{\w}{f(\boldx)f(\boldx^*)^T} = \phi(\boldx)^T\ExpectationWRT{\w}{\w\w^T}\phi(\boldx^*) = \phi(\boldx)^T\wpriorcov\phi(\boldx^*)$ which indicates that $f(\boldx)$ and $f(\boldx^*)$ are jointly Gaussian with mean $0$ and covariance $\phi(\boldx)^T\wpriorcov\phi(\boldx^*)$. More generally in GP context, the covariance matrices $K$ are constructed by a symmetric covariance kernel function $\kernelfunction{\boldx_i}{\boldx_j}$ in which the entries of $K$ (also called \textit{Gram matrix}) are defined by $K_{ij}=\kernelfunction{\boldx_i}{\boldx_j}$. The key component of GP is the choice of the kernel which determines the smoothness and characteristics of the underlying functions, and the hyperparameters associated with the kernel that control its behavior. In this study, we will apply the widely-used \textit{squared exponential} kernel defined by:

\begin{equation}\label{Equation::squared_exponential_kernel_definition}
    k(\boldx_i, \boldx_j) = \exp\left(-\frac{|\boldx_i-\boldx_j|^2}{2l^2}\right),
\end{equation}
where $l$ is the \textit{characteristic length scale} which controls how sensitive the kernel is to small changes in the explanatory variables. When $l$ is small nearby data points will have a high covariance, which decreases rapidly as data points move away from each other. In this case, the GP model is more flexible and can capture fine-grained variations in the data. On the other hand, when the length scale is large, the kernel function is less sensitive to small changes in the explanatory variables resulting in a smoother kernel function, and data points farther apart will still have a relatively high covariance. In this situation, the GP model is less flexible and captures more general trends in the data, filtering out noise better. For a list of other common GP kernels see e.g. \citep[][Ch. 4]{Rasmussen2005}.

\subsection{Horvitz-Thompson and difference estimator}\label{Section::Howrwitz_and_Difference_estimator}

As is well-known in the literature of survey sampling, an unbiased estimator for population total $t_y=\sumPopulation y_k$ is the \textit{Horvitz-Thompson}  estimator (known also sometimes as the $\pi$-estimator) \citep{Fuller2009}: 

\begin{equation}
    \horvitzThompsonEstimator = \sumSample\frac{y_k}{\pi_k},
\end{equation}
which has a variance: 

\begin{equation}
    \Vsampledesign{\horvitzThompsonEstimator} = \sumPopulation\sumPopulation (\pi_{kj}-\pi_k\pi_j)\frac{y_k}{\pi_k}\frac{y_j}{\pi_j}.
\end{equation}
Another unbiased estimator for the total $t_y$, similar to the $\pi$-estimator, is the \textit{difference estimator} \citep[DE, ][]{model2003sarndal} defined as: 

\begin{equation} \label{difference_estmator_definition}
    \differenceEstimator = \sumPopulation \hat{y}_k + \sumSample \frac{D_k}{\pi_k},
\end{equation}
where $\hat{y}_k$ is model (e.g. linear regression) produced estimate for the response value of $y_k$ with explanatory variables $\boldx_k$ and $D_k=y_k-\hat{y}_k$. The difference estimator has a variance:

\begin{equation}\label{Equation::difference_estimator_variance}
    \Vsampledesign{\differenceEstimator} = \sumPopulation\sumPopulation (\pi_{kj}-\pi_k \pi_j)\frac{D_k}{\pi_k}\frac{D_j}{\pi_j},
\end{equation}
from which we notice that if the model produces good estimates for the population response variables $\hat{y}_k$, then the difference terms $D_k$, i.e. residuals, will be small, and hence the variance will be potentially much smaller than that of the $\pi$-estimator. In general, we can evaluate the performance of an estimator $B$ relative to some reference baseline estimator $A$ using the ratio of their variances called \textit{relative efficiency}. In our case, if we can produce a good surrogate model for the response population we will have $\Vsampledesign{\differenceEstimator} / \Vsampledesign{\horvitzThompsonEstimator} < 1$ indicating we have more efficient estimator than the $\pi$-estimator. More details and key derivations of DE is shown in the Appendix.

\subsection{Bayesian optimization}\label{Section::Bayesian_optimization}
Bayesian optimization \citep[BO, e.g.][]{Archetti2019,garnett_2023} is a powerful and efficient technique used to optimize complex and expensive-to-evaluate functions. It finds its applications in various fields, including machine learning, engineering, and scientific research \citep[e.g.,][]{Remi2018,VANNIEKERK2022126,Kang2023,NIPS2012_05311655}. At its core, Bayesian optimization combines probabilistic modeling with optimization methods to intelligently explore the parameter space of a given function, aiming to discover the optimal set of parameters that maximize or minimize the function's output. By iteratively selecting the next set of parameters to evaluate based on a probabilistic surrogate model, BO strikes a balance between exploration and exploitation, allowing it to efficiently converge to the global optimum while minimizing the number of function evaluations. This approach is particularly useful when each evaluation of the function is time-consuming, expensive, or resource-intensive, making it a valuable tool for tackling real-world optimization challenges.

\subsubsection{Acquisition functions}\label{Section: Acquisition functions}

The essence of BO is focused on the general optimization problem of either minimizing or maximizing an \textit{objective function} $\objectiveFunction{\boldx}$ using an \textit{acquisition function} $\acquisitionFunction{\boldx}: \mathbb{R}^\xdimsymbol \to \mathbb{R}$ used to guide the selection of next point to evaluate in the optimization process of $\objectiveFunction{\boldx}$. In this work, we assume the acquisition function is always positive, i.e. $\acquisitionFunction{\boldx} > 0\; \forall \;\boldx\in\xpopulation$. Moreover, we interpret a higher acquisition value at a new point as an indication of higher significance or utility for that particular point w.r.t to optimizing $\objectiveFunction{\boldx}$. We will next go through the acquisition functions used in this study, which are all summarized in Table \ref{Table::Acquisition_functions}.

\subsubsection*{Predictive uncertainty}
As the first acquisition function we use the \textit{predictive uncertainty} $\predictiveUncertainty{\boldx}$ which we define simply as the standard deviation $\sigma(\boldx)$ of the posterior predictive distribution of the response population variable $y$ under consideration. The rationale for employing the standard deviation of the response posterior predictive distribution as the acquisition function is in the utilization of the surrogate model's uncertainty at those specific locations. The idea is that identifying points with heightened predictive uncertainty in the surrogate model could yield valuable insights. This heightened uncertainty is anticipated in regions of the input space where e.g. the explanatory variables significantly deviate from the data already observed in the data set $\dataset$, thus potentially offering novel and informative data points for model improvement. In other words, the goal for $\predictiveUncertainty{\boldx}$ is to select a new point which maximizes the objective function $\objectiveFunction{\boldx} := \predictiveUncertainty{\boldx} = \sigma(\boldx)$. Note that in this special case, the objective and the acquisition functions are equal because the defined goal here is to maximize variance in subsequent steps. In other words, if $f(\boldx)|\dataset \sim \NormalDistribution{\mu(\boldx)}{\sigma^2(\boldx)}$ is the posterior predictive surrogate model for the response $y$, then the predictive uncertainty guides to select new point $\boldx^*$ such that $\boldx^* = \argmax_{\boldx} \sigma(\boldx)$.

\subsubsection*{Inverted lower confidence bound}\label{Section::ILCB_definition}
In our second acquisition function case, we first define the objective function as the change in surrogate model's out-of-sample performance, in terms of \textit{mean absolute error} (MAE) for the response $y$, when data point $\datapoint$ is included into data set $\dataset$ used to construct $f$. We denote this as $\objectiveFunction{\boldx} := \Delta  \text{MAE}^f(\boldx) = \text{MAE}_{\dataset \cup \datapoint}^f-\text{MAE}_{\dataset}^f$ in which $f$ denotes a surrogate model for the response $y$ and $\datapoint=(\boldx, y)$. Note that the inclusion of a data point $\datapoint$ improves the model $f$ when $\text{MAE}_{\dataset \cup \datapoint}^f <\text{MAE}_{\dataset}^f$. Our goal is thus to minimize the objective function $\objectiveFunction{\boldx}$, focusing on data points that contribute most to enhancing the surrogate model $f$. In our case study in section \ref{Section::case_study}, we use Monte Carlo simulation to estimate the objective function $\objectiveFunction{\boldx}$ with empirical data.

We now define the \textit{inverted lower confidence bound} (ILCB) acquisition function as $\InvertedlowerConfidenceBound{\boldx} = -\lowerConfidenceBound{\boldx} = \lambda\sigma(\boldx) -\mu(\boldx)$ \citep[see e.g.,][]{Archetti2019}, where $\lambda\geq0$, and $\mu(\boldx), \sigma(\boldx)$ are the mean and standard deviation of a surrogate model $h$ built for the objective $\objectiveFunction{\boldx}$ subject to minimization. The goal with this acquisition function is to select new point which maximizes ILCB and hence, with highest expectation, minimizes the objective $\objectiveFunction{\boldx}$. The rationale behind the definition of ILCB is to balance between the exploitation $\mu(\boldx)$ and exploration $\sigma(\boldx)$. In other words, the ILCB gives highest utility for points which on expectation, based on the observed data $\dataset$, resulted in small objective values, as well as which have high uncertainty in the performance change. If we use GP as a surrogate model, high uncertainty is given for new points which differ greatly in explanatory features from the observed data $\dataset$. Also, the greater the $\lambda$ value, the more weight is given to exploration instead of exploitation, i.e., we are willing to take more risks when exploring the sample points. In this work we parameterized the exploration as $\lambda = 0.2$, meaning that ILCB favors points which on expectation result in high model performance improvement regardless of the risk involved.

\subsubsection*{Expected improvement}
In this and the following section, our objective function will be the same as in the ILCB case. The \textit{expected improvement} \citep[EI,][]{Mockus1978} is based on the \textit{improvement function} $\Improvement{\boldx} \coloneqq \max\{\fmin-\objectiveFunction{\boldx}, 0\}$ where $\fmin$ is the current minimum observed value of the objective function $\objectiveFunctionSymbol$. Regarding the problem of minimizing $\objectiveFunctionSymbol$, in EI we want to select new point which has a highest expected improvement in that point. The EI is defined as:  

\begin{equation}
    \ExpectedImprovement{\boldx} = \ExpectationGeneral{\objectiveFunctionSymbol|\dataset}{\Improvement{\boldx}}=\sigma(\boldx)\left[z\Psi\left(z\right) + \psi\left(z\right)\right],
\end{equation}
where the expectation is taken with respect to the posterior predictive distribution of the objective function $\objectiveFunctionSymbol$. Also, we have $z=(\fmin-\mu(\boldx))/\sigma(\boldx)$, where $\mu(\boldx), \sigma(\boldx)$ are the mean and standard deviation of the surrogate posterior predictive model built for $\objectiveFunctionSymbol$, and $\Psi$ and $\psi$ denote the cumulative distribution and probability density functions of a standard normal distribution $\mathcal{N}\left(0, 1\right)$ respectively. The idea of EI is simple, select new data point which is expected to have the highest objective improvement, based on the observed data so far. Derivation of the EI and its variance is shown in the Appendix of this study.

\subsubsection*{Scaled expected improvement}
Lastly, we have the \textit{scaled expected improvement} \citep[SEI,][]{noe2018new} acquisition function which not only takes into account the amount of expected improvement, but also the uncertainty (i.e. the variance) of the improvement $\Improvement{\boldx}$. The variance of the improvement function is defined as: 

\begin{equation}
\begin{split}
    \VarianceGeneral{\objectiveFunctionSymbol|\dataset}{\text{I}(\boldx)} &= \ExpectationGeneral{\objectiveFunctionSymbol|\dataset}{\Improvement{\boldx}^2}-(\text{EI}(\boldx))^2 \\
    &= \sigma^2(\boldx)\{(z^2+1)\Psi(z)+z\psi(z)\} - (\text{EI}(\boldx))^2,
\end{split}
\end{equation}
which is used to define the SEI, that is: 
\begin{equation}
    \ScaledExpectedImprovement{\boldx} = \frac{\ExpectedImprovement{\boldx}}{\sqrt{\VarianceGeneral{\objectiveFunctionSymbol|\dataset}{\text{I}(\boldx)}}}. 
\end{equation}
In other words, regarding the minimization of the objective function $\objectiveFunctionSymbol$, the SEI selects new points which on expectation result in highest improvement with highest certainty.

\begin{table}[t]
\centering
\renewcommand{\arraystretch}{1.8} 
\begin{tabular}{|>{\centering\arraybackslash}m{0.4\textwidth}|>{\centering\arraybackslash}m{0.1\textwidth}|>{\centering\arraybackslash}m{0.35\textwidth}|>{\centering\arraybackslash}m{0.2\textwidth}|}
\hline
\rowcolor{lightgray}
\textbf{Acquisition function} & \textbf{Notation} & \textbf{Definition}  \\
\hline
predictive uncertainty & $\predictiveUncertainty{\boldx}$ & $\sigma(\boldx)$  \\
\hline
inverted lower confidence bound & $\text{ILCB}(\boldx)$ & $\lambda\sigma(\boldx)-\mu(\boldx)$  \\
\hline
expected improvement & $\text{EI}(\boldx)$ & $\sigma(\boldx)\left[z\Psi\left(z\right) + \psi\left(z\right)\right]$  \\
\hline
scaled expected improvement & $\text{SEI}(\boldx)$ & $\frac{\sigma(\boldx)\left[z\Psi\left(z\right) + \psi\left(z\right)\right]}{\sqrt{\sigma^2(\boldx)\{(z^2+1)\Psi(z)+z\psi(z)\} - (\text{EI}(\boldx))^2}}$  \\
\hline
\end{tabular}
\caption{List of acquisition functions used in this work.}
\label{Table::Acquisition_functions}
\end{table}

\section{Model-assisted sampling with Bayesian optimization}
In this section we present a novel model-assisted sampling design based on the difference estimator of section \ref{Section::Howrwitz_and_Difference_estimator} and Bayesian optimization of section \ref{Section::Bayesian_optimization}. Our proposed method is based on the following observation: We know that difference estimator $\differenceEstimator$ in \ref{difference_estmator_definition} is an unbiased estimator of the target population total, and the variance of this estimator can potentially be much more efficient than that of the $\pi$-estimator by minimizing the residual terms $D_k$ in \ref{Equation::difference_estimator_variance}, given that a suitable surrogate model is used. Therefore, there is a strong motivation to utilize an effective surrogate model, as its performance directly influences the outcome. Consequently, we aim to develop an unbiased sampling procedure that not only considers optimizing the surrogate model but also has the potential to yield a lower variance difference estimator by strategically minimizing the residuals $D_k$. Given the previous discussion, we define the main steps of our proposed method: 
\begin{enumerate}
    \item Assume a prior observed data set $\datasetPrior = (\xPrior, \yPrior)$ (e.g. from earlier surveys) and a sampling population $\population$. Use BO to estimate an acquisition function $\acquisitionFunction{\boldx}$ for all $\boldx \in\xpopulation$ using $\datasetPrior$ and the surrogate model. In essence, the acquisition function gauges the inclusion utility of each potential sample point $(\boldx, y) \in \population$. In other words, we let past accumulated knowledge to optimize future actions.
    \item  Given the acquisition function $\acquisitionFunction{\boldx}$ for all sample points in $\population$, we construct the sampling design by defining the population inclusion probabilities as functions of the acquisition values $\pi(\acquisitionFunction{\boldx})$. The intuition is, the higher the acquisition value (i.e. utility), the higher the inclusion probability. 
    \item Obtain a probability sample using the constructed sampling design, estimate a surrogate model for population target variable and construct unbiased population estimation correspondingly.
\end{enumerate}
For more explicit definition, we present the pseudocode of our proposed method in Algorithm \ref{alg:BO_sampling}.
\begin{algorithm}
\caption{Sampling design construction and population paremeter estimation with BO.}
\label{alg:BO_sampling}
\begin{algorithmic}[1]
    \REQUIRE Prior data set $\datasetPrior$, surrogate probabilistic model $f$, sampling population $\population$.
    \STATE Estimate positive acquisition function values $\acquisitionFunction{\boldx}$ for each $(\boldx, y)\in\population$ using $\datasetPrior$ and BO.
    \STATE Form the sampling design by defining the inclusion probabilities using min-max normalization as: 
    \begin{equation} \label{sampling_pseudocode_definition}
    \pi(\acquisitionFunction{\boldx}) =
\begin{cases}
    \frac{\pseudocodeSymbol{\boldx}-\pseudocodeSymbolMin}{\pseudocodeSymbolMax-\pseudocodeSymbolMin}, & \text{if } \pseudocodeSymbolMin
    < \pseudocodeSymbol{\boldx} < \pseudocodeSymbolMax \\
    1-\epsilon, & \text{if } \pseudocodeSymbol{\boldx} = \pseudocodeSymbolMax \\
    \epsilon, & \text{if } \pseudocodeSymbol{\boldx} = \pseudocodeSymbolMin,
\end{cases}
\end{equation}
where $\pseudocodeSymbolMin, \pseudocodeSymbolMax$ are the minimum and maximum acquisition values for the sample points in $\population$ respectively, and $\epsilon>0$ is a small positive number to guarantee that $0 < \pi(\acquisitionFunction{\boldx}) < 1$ always.
    \STATE Produce a probability sample $\sample$ using the model optimized sampling design and set $\datasetPosterior = \datasetPrior \cup \sample$.
    \STATE Fit a surrogate model $f$ using $\datasetPosterior$ and use it to estimate rest of the population and corresponding parameters respectively.
\end{algorithmic}
\end{algorithm}

\section{Case study}\label{Section::case_study}
In this section we conduct an empirical experiment with the proposed method against a baseline sampling design, simple random sampling (SRS). Our empirical analysis implements the steps of Algorithm \ref{alg:BO_sampling} in 10,000 independent Monte Carlo sampling simulations in which we estimate three population parameters of interest: response total, mean and distribution (i.e. the response histogram). In the beginning of each simulation, a random prior probability SRS sample $\datasetPrior$ is taken as required by the method. The same prior data per sample simulation is used for all compared sampling designs to guarantee equal starting states. Our empirical data population consisted in total of 1920 observations covering tree volume and laser scanning data which we cover in more detail in sections \ref{Section::response_data} and \ref{Section::explanatory_data}. The random prior sample $\datasetPrior$ comprised of 100 observations, accounting for approximately 5.2\% of the empirical population dataset. For the sampling simulations, a sample size of 50 observations was employed, constituting approximately 2.7\% of the remaining empirical population, which consisted of 1820 observations correspondingly.

\subsection{Response variable: field plot level tree volume}\label{Section::response_data}
The used response field data measured in a study area located in central Finland, covering approximately 5800 km$^2$, describes the overall plot level tree volume (m$^3$/ha). The study area is mainly covered by forestry land on moraine and peat soils belonging to middle boreal vegetation zone. The topography of the area is relatively flat with elevation ranging from 100 to 200 m above sea level. The main tree species are Scots pine, Norway spruce, silver birch and downy birch.
The field observations have been sampled based on systematic cluster sampling, where the clusters were inverted L-shaped, with full cluster having 8 sample plots with a spacing of 250 m.  The horizontal and vertical distance between the clusters was 4.3 km \citep{Tomppo2016}. The field plots circular sample plots with radius of 9 m. All trees whose diameter at breast height was 4.5 cm or more were measured as tally trees from the plots. Plot level volumes were calculated on the basis of tally trees using species specific volume functions by \citep{Laasasenaho1982}.

\subsection{Explanatory variables: laser scanning data}\label{Section::explanatory_data}
Airborne laser scanning (ALS) data were acquired originally for the Finnish Forest Centre in June–August 2013 using an Optech Gemini ALTM laser scanner. The scanning altitude was 1730 m with a maximum zenith angle of 20° and side overlap of 20\%. The average pulse density of the ALS point cloud was 0.9 pts/m$^2$, scanning strip overlap was 20\%, half scan angle 20 degrees. Pulse returns from higher than 1.3 m above the ground were classified as canopy returns. More extensive description of the ALS data is presented by \citep{Tomppo2016}. 
The ALS features applied in this study were extracted from 16x16 m square window centred around the field sample plot centre points. The following features (in total 16) extracted from the ALS point cloud data were applied \citep{Naesset2002,Maltamo2009,Tomppo2016}: 
1.	Average value of height above ground (H) for canopy returns (m) 
2. H at which the following percentiles (0\%, 5\%, 10\%, 20\%, ..., 85\%, 90\%, 95\%, 100\%)  of first canopy returns pulses were  accumulated (m) 
3.	Proportion of canopy returns relative to all points (\%).

\subsection{Performance metrics}

We will evaluate the performance of the sampling designs  of the study with four metrics (all having optimum value at $0$) against the SRS baseline:
\begin{enumerate}
    \item Absolute difference between true and model estimated population mean $\left|\ExpectationSymbol{y}-\ExpectationSymbol{\hat{y}}\right|$.
    \item Kullback-Leibler \citep[KL,][]{Kullback_1951} divergence between true $(P)$ and estimated $(Q)$ population distribution $D_{\text{KL}}(P \| Q) = \sum_{y\in\population} P(y) \log\left(\frac{P(y)}{Q(y)}\right)$.
    \item Absolute difference between true and DE-estimated population total $\left|t_y-\differenceEstimator\right|$.
    \item Absolute difference between true and DE-estimated population total $\left|t_y-\differenceEstimator\right|$ with normalized inclusion probabilities $\pi_i^{*} = \frac{\pi_i}{\sum_{\population}\pi_k}$ so that $\sum_{\population}\pi_i^{*} = 1$.
\end{enumerate}
To give rationale for metric (4), we first notice in the difference estimator $\differenceEstimator$ in \ref{difference_estmator_definition} that individual data points having small inclusion probabilities $\pi_k$ will effectively have a large weight in the total estimation. Because we define the inclusion probabilities for all data points in SRS design being equal to $\pi_k=1/N$ where $N$ is the size of the population $\population$, it can occur that the difference estimator for SRS overestimates the population total in metric (3) simply because of the small inclusion probabilities. To reduce the effect of this and make it more comparable to the proposed sampling designs in this study, we normalize the inclusion probabilities in metric (4) to sum to $1$ for all sampling designs. Furthermore, we will evaluate the statistical significance of our results using the non-parametric Mann–Whitney U test \citep{MannWhitney}.

\subsection{Results}
\begin{figure}[H]
    \centering
    \begin{minipage}{\textwidth}
        \centering
        \begin{minipage}[b]{0.49\textwidth}
            \centering
            \includegraphics[width=\linewidth]{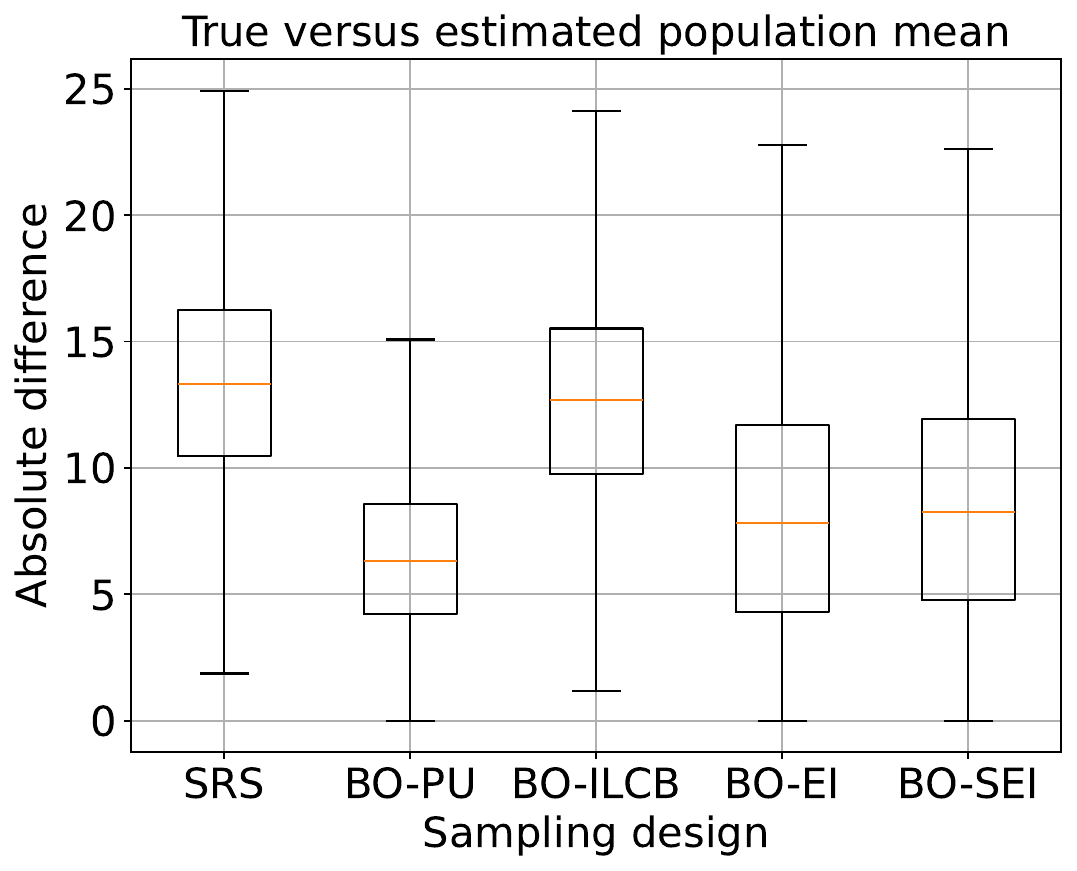}
        \end{minipage}
        \hfill
        \begin{minipage}[b]{0.49\textwidth}
            \centering
            \includegraphics[width=\linewidth]{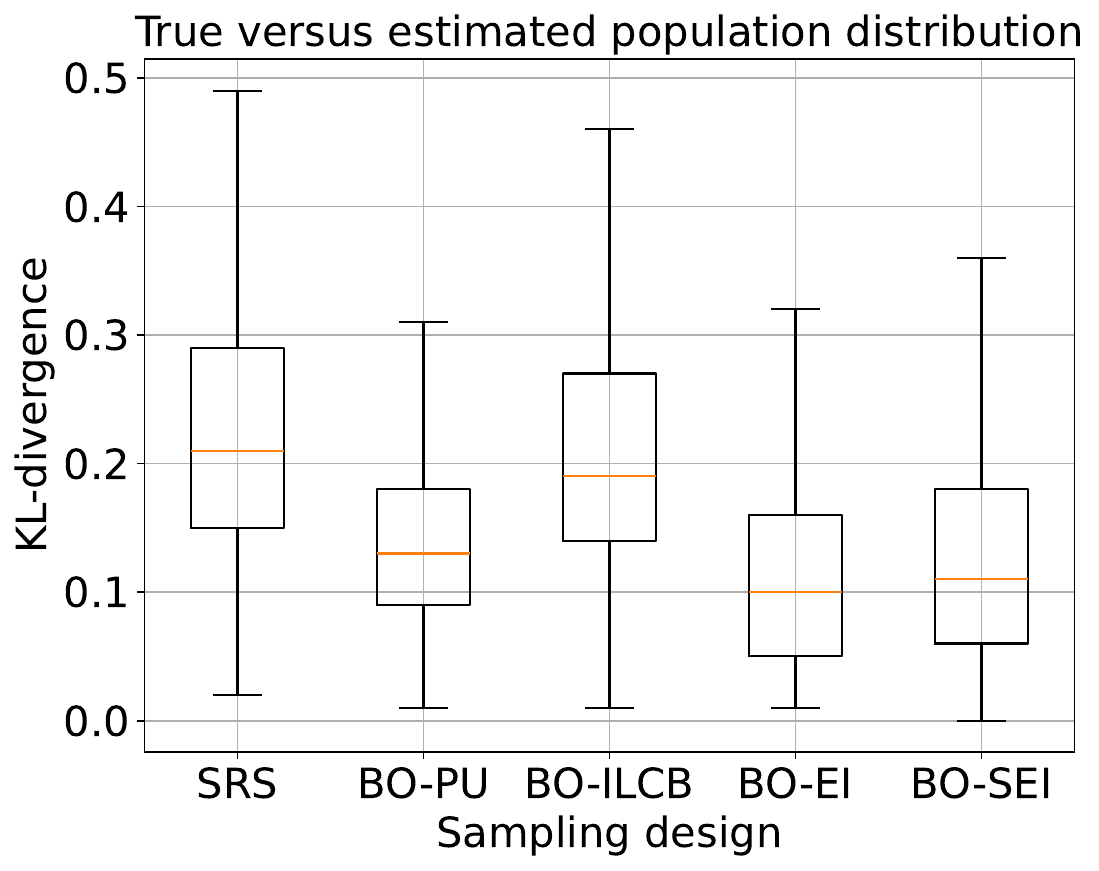}
        \end{minipage}
        \vspace{10pt}
        \begin{minipage}[b]{0.49\textwidth}
            \centering
            \includegraphics[width=\linewidth]{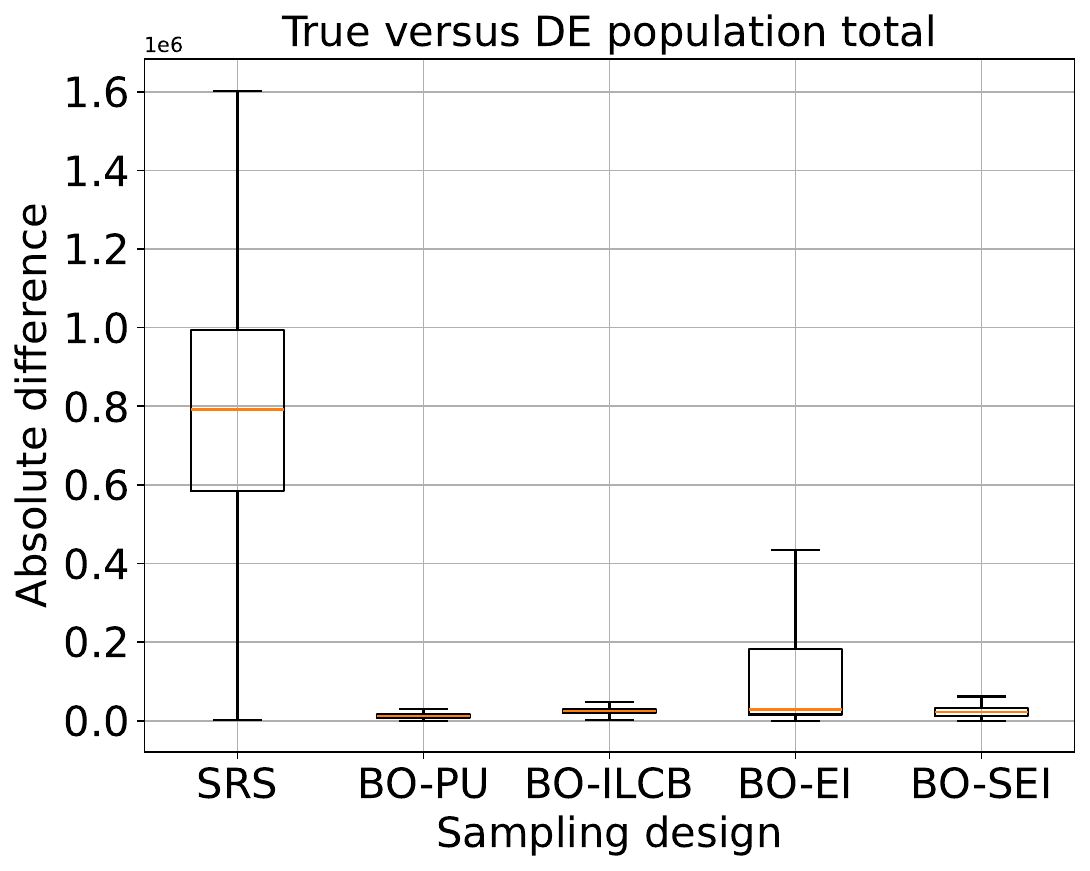}
        \end{minipage}
        \hfill
        \begin{minipage}[b]{0.49\textwidth}
            \centering
            \includegraphics[width=\linewidth]{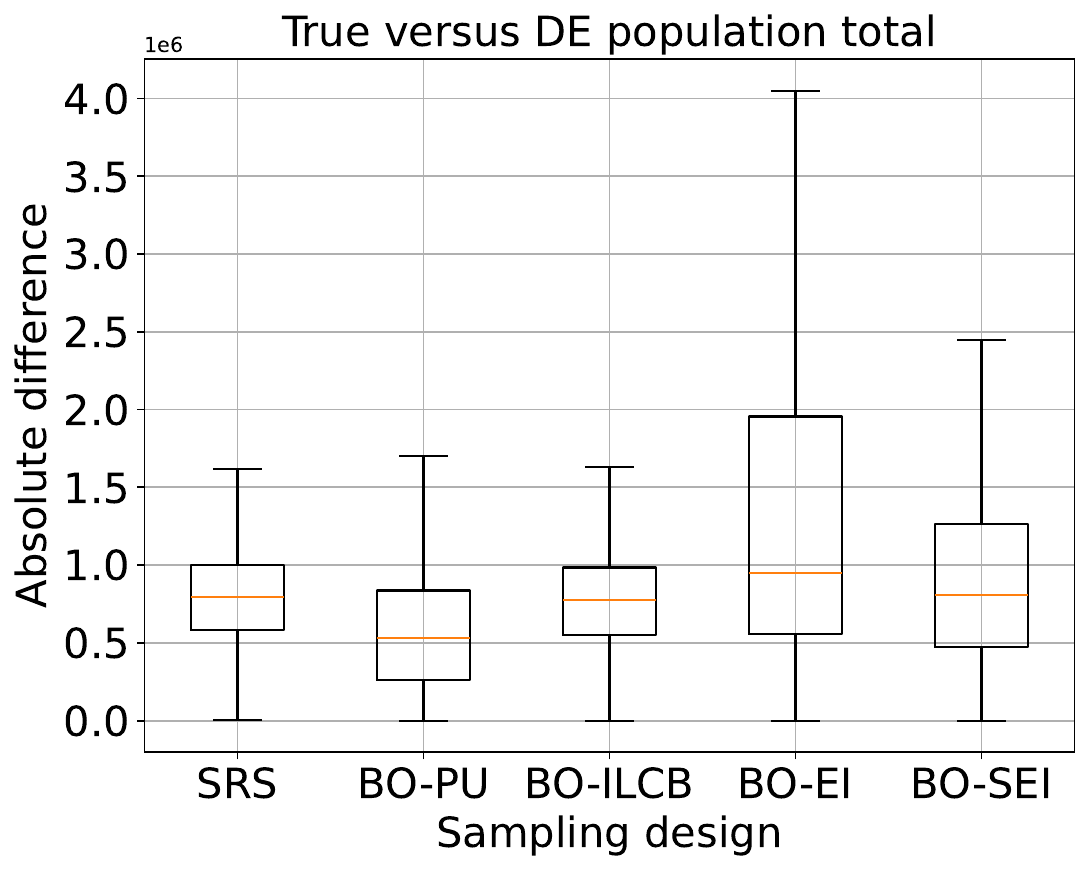}
        \end{minipage}
        \caption{Empirical sampling simulation results with 10,000 repeats for different sampling methods (x-axis) with boxplots for respective metrics. Median of the simulation results is represented with a solid orange line. Top-left: Absolute difference between true and estimated response population mean. Top-right: KL-divergence between the true and estimated response population distribution. Bottom-left: Absolute difference between true response total and the difference estimator $\differenceEstimator$ with unnormalized inclusion probabilities. Bottom-left: Absolute difference between true response total and the difference estimator $\differenceEstimator$ with normalized inclusion probabilities.}
        \label{Figure:study_results}
    \end{minipage}
\end{figure}

We have illustrated the empirical study results in Figure \ref{Figure:study_results} with boxplots. The proposed method is generally denoted with the prefix "BO-" with the latter part depending on what acquisition function from section \ref{Section: Acquisition functions} was used. The baseline method SRS is always the leftmost boxplot. Also, the corresponding Mann-Whitney U test p-values with the alternative hypothesis $H_1$ that BO sampling methods produce stochastically smaller corresponding metric values than SRS is presented in Table \ref{table:mannwhitneyu_results}. We notice from the results that in all cases, besides in metric (4) for BO-EI and BO-SEI, all results are statistically significant. This shows that the proposed sampling method outperforms the baseline method in probability sampling, especially with the BO-PU sampling design. We also notice the significant difference in population total estimation using the DE if inclusion probabilities are normalized or not. The BO-EI and BO-SEI sampling designs have similar performances in all of the four metrics with BO-EI having larger variability than BO-SEI. Overall, the BO-ILCB produces worst results of the four tested acquisition functions and not showing great advantage over the SRS. However, the $\lambda$-parameter (exploration) of the ILCB was fixed to $\lambda=0.2$, meaning low exploration in the BO-ILCB, and could be a subject of optimization. Additionally, BO-PU, BO-EI, and BO-SEI exhibit positively skewed distributions in the performance metrics

\begin{table}[H]
\centering
\caption{Mann–Whitney U test p-values with the alternative hypothesis $H_1$ that the BO sampling methods produce stochastically smaller corresponding metric values than SRS.}
\label{table:mannwhitneyu_results}
\begin{tabular}{|l|c|c|c|c|}
\hline
\rowcolor[HTML]{EFEFEF}
\textbf{Metric} & \cellcolor[HTML]{EFEFEF}\textbf{BO-PU} & \cellcolor[HTML]{EFEFEF}\textbf{BO-ILCB} & \cellcolor[HTML]{EFEFEF}\textbf{BO-EI} & \cellcolor[HTML]{EFEFEF}\textbf{BO-SEI} \\
\hline
Mean difference & 0 & 3.50e-30 & 0 & 0 \\
\hline
KL-divergence & 0 & 9.81e-34 & 0 & 0 \\
\hline
Population total difference & 0 & 0 & 0 & 0 \\
\hline
Population total difference (norm. $\pi_i$) & 0 & 1.84e-09 & 1 & 1 \\
\hline
\end{tabular}
\end{table}

\section{Discussion}

Sampling design optimization has been extensively researched, with studies focusing on methods such as leveraging auxiliary data to improve sampling efficiency \citep[e.g.,][]{Saarela2015,Singh2003}. It is widely recognized that regression estimators can enhance population parameter estimation by incorporating auxiliary data \citep{model2003sarndal}. Additionally, existing literature has proposed novel approaches utilizing auxiliary data to achieve balanced sampling, as demonstrated by the local pivotal method \citep{Grafstrom2014}. However, to our knowledge, the application of Bayesian optimization (BO) within the context of survey sampling remains mostly unexplored in the existing literature. While numerous studies have investigated various methodologies for survey design and sampling techniques, the incorporation of BO into this domain presents a novel opportunity for advancement.

The empirical results of this work indicate overall best results for the BO-PU over all tested sampling designs, with the exception on the KL-divergence metric. This suggests that especially the surrogate model's uncertainty on the response variable is an improving factor in contributing to model-assisted sampling design optimization. In recent study by \citep{Wang2021NonuniformNS} a similar discovery was made in a binary classification model context with an empirical data set of size 0.3 trillion observations. In this study, the authors demonstrated and proved the effectiveness of a strategic negative sampling approach, emphasizing the value of selecting samples that induce uncertainty for the model.

A noteworthy observation concerning the diference estimator outlined in Equation \ref{difference_estmator_definition} and the proposed sampling methodology is that instances where the model exhibits high confidence result in exceedingly small inclusion probabilities, denoted as $\pi_k$. Consequently, this imparts substantial weight to observations characterized by low uncertainty during estimation. However, it is reasonable to presume that the model excels in accurately estimating observations with minimal uncertainty, thereby yielding low difference terms ($D_k$) and subsequently mitigating the impact of the elevated weighting. Furthermore, it is imperative to highlight that in the proposed methodology, if the BO sampling approach confers nearly uniform inclusion probabilities across all population observations, each observation is assigned equal weights, aligning the behavior of the proposed method with that of the baseline SRS.

Moreover, within the scope of this work, a GP surrogate model was employed in the BO sampling methodology. It is crucial to note that the sampling method described is versatile and remains applicable with any surrogate model that facilitates the computation of the respective acquisition functions. This flexibility allows for the integration of alternative surrogate models based on the specific requirements of the study or available data.

\section{Conclusions}
This study introduced the application of Bayesian optimization in the context of model-assisted sampling. Our methodological framework rests on the premise that utilizing a well-suited surrogate model for the particular study and data set can potentially yield a more efficient estimator compared to the Horvitz-Thompson estimator via the optimization of the sampling design. Empirical results support this proposition, revealing statistically significant results that underscore the effectiveness of our proposed method, particularly when incorporating the model's uncertainty in the optimization of sampling design.

\bibliographystyle{abbrvnat}
\bibliography{sample}

\newpage

\section*{Appendix}
In this appendix we show the derivations of some of the key methods used in this study.
\subsection*{The population total and difference estimator}

A common population parameter to be estimated is the population total $t_y$ of the response variable $y$, that is:
\begin{equation}
    t_y = \sum_\population y_k,
\end{equation}
where $y_k$ is the $k$th observation in the population $\population$. The total can be rewritten in the following way:
\begin{equation} \label{difference_total}
\begin{split}
t_y &= \sum_\population y_k \\
 &= \sum_\population \hat{y}_k + \sum_\population\left(y_k -\hat{y}_k\right) \\
 &= \sum_\population \hat{y}_k + \sum_\population D_k,
\end{split}
\end{equation}
where $\hat{y}_k$ is a model estimated value of $y_k$ and $D_k=y_k-\hat{y}_k$. For estimating the population total we can use the difference estimator (DE) defined as \citep{model2003sarndal}: 
\begin{equation}
    \differenceEstimator = \sum_\population \hat{y}_k + \sum_s \frac{D_k}{\pi_k}.
\end{equation}
The difference estimator is an unbiased estimator of the total as can easily be shown. Let us denote $\mathbb{E}_{p(s)}[\cdot]$ as expectation w.r.t all possible samples and $I_k$ as an indicator random variable taking values $1$ or $0$ if observation $y_k$ was or was not selected in the sample $\sample$ respectively. It then follows that: 
\begin{equation} \label{difference_estimator_proof}
\begin{split}
\mathbb{E}_{p(s)}\left[\differenceEstimator\right] &= \mathbb{E}_{p(s)}\left[\sum_\population \hat{y}_k + \sum_s \frac{D_k}{\pi_k}\right] \\
 &= \sum_\population \hat{y}_k + \mathbb{E}_{p(s)}\left[\sum_\population I_k\frac{D_k}{\pi_k}\right] \\
 &= \sum_\population \hat{y}_k + \sum_\population \mathbb{E}_{p(s)}\left[I_k\right]\frac{D_k}{\pi_k} = \sum_\population \hat{y}_k + \sum_\population \pi_k\frac{D_k}{\pi_k}\\
 &= \sum_\population \hat{y}_k + \sum_\population D_k = t_y.\;\;\qed
\end{split}
\end{equation}

Next, we introduce additional notation. The inclusion probability that both observations $y_k$ and $y_i$ are selected in the sample is denoted as $\pi_{kj}=\mathbb{E}_{p(s)}\left[I_kI_j\right]$. The covariance between random variables $I_k, I_j$ is denoted as $\Delta_{kj}=\text{Cov}\left[I_k, I_j\right]=\pi_{kj}-\pi_k \pi_j$, and also we define $\hat{D}_k=D_k/\pi_k$. With these we derive the variance of $\differenceEstimator$:

\begin{equation} \label{difference_estimator_variance}
\begin{split}
\mathbb{V}\left[\differenceEstimator \right] &= \Esampledesign{(\differenceEstimator-t_y)^2} = \Esampledesign{\differenceEstimator^2}-t_y^2\\
 &= \Esampledesign{\left(\sum_\population\hat{y}_k\right)^2 + 2\sum_\population\hat{y}_k \sum_\population I_k\frac{D_k}{\pi_k} + \left(\sum_\population I_k\frac{D_k}{\pi_k}\right)^2} -t_y^2\\
 &= \left(\sum_\population\hat{y}_k\right)^2 + 2\sum_\population\hat{y}_k \sum_\population (y_k-\hat{y}_k) + \Esampledesign{\left(\sum_\population I_k\frac{D_k}{\pi_k}\right)^2} -t_y^2\\
 &= 2\sum_\population\sum_\population\hat{y}_k  y_k - \left(\sum_\population\hat{y}_k\right)^2 + \Esampledesign{\left(\sum_\population I_k\frac{D_k}{\pi_k}\right)^2} -t_y^2\\
  &= 2\sum_\population\sum_\population\hat{y}_k y_j - \sum_\population\sum_\population\hat{y}_k\hat{y}_j + \sum_\population\sum_\population \Esampledesign{I_k I_j}\frac{D_k D_j}{\pi_k \pi_j} -t_y^2\\
  &=  \sum_\population\sum_\population \pi_{kj}\frac{D_k D_j}{\pi_k \pi_j}  +2\sum_\population\sum_\population\hat{y}_k y_j - \sum_\population\sum_\population\hat{y}_k\hat{y}_j - \sum_\population\sum_\population y_ky_j \\
  &=  \sum_\population\sum_\population \pi_{kj}\frac{D_k D_j}{\pi_k \pi_j}   - \sum_\population \sum_\population D_k D_j \\
  &=  \sum_\population\sum_\population (\pi_{kj}-\pi_k \pi_j)\frac{D_k D_j}{\pi_k \pi_j} = \sum_\population\sum_\population \Delta_{kj}\hat{D}_k\hat{D}_j.\;\;\qed
\end{split}
\end{equation}
An unbiased estimator for the DE variance in \ref{difference_estimator_variance} is:

\begin{equation} \label{difference_estimator_variance_estimator}
    \hat{\mathbb{V}}\left[\differenceEstimator\right] = \sum_s\sum_s \hat{\Delta}_{kj}\hat{D}_k\hat{D}_j,
\end{equation}
where $\hat{\Delta}_{kj}=\Delta_{kj}/\pi_{kj}$. This can be easily proved: 

\begin{equation} \label{difference_estimator_variance_estimator_proof}
\begin{split}
\Esampledesign{\hat{\mathbb{V}}\left[\differenceEstimator\right]} &= \Esampledesign{\sum_s\sum_s \hat{\Delta}_{kj}\hat{D}_k\hat{D}_j} \\ &= \Esampledesign{\sum_\population\sum_\population I_kI_j\hat{\Delta}_{kj}\hat{D}_k\hat{D}_j}\\
&= \sum_\population\sum_\population \Esampledesign{I_kI_j}\hat{\Delta}_{kj}\hat{D}_k\hat{D}_j \\
&= \sum_\population\sum_\population \pi_{kj}\hat{\Delta}_{kj}\hat{D}_k\hat{D}_j \\
&= \sum_\population\sum_\population \Delta_{kj}\hat{D}_k\hat{D}_j = \mathbb{V}\left[\differenceEstimator \right].\;\;\qed
\end{split}
\end{equation}
As we have discussed in this work, notice that we can decrease the DE estimator variance by having small $D_k$ values for all $k$.

\subsection*{Expected improvement}\label{Section::appendix_expected_improvement}
In this section we derive the expected improvement (EI) and its variance following the work of \citep{noe2018new}. We assume our objective is to minimize a given objective function $\objectiveFunction{\boldx}$ for which we have available a surrogate posterior predictive model $p\left(\objectiveFunctionSymbol | \dataset\right)$ based on past observations $\dataset$ from the unknown function $\objectiveFunction{\boldx}$. All expectations presented below are taken w.r.t to random variable $g$ with conditional posterior predictive distribution $p\left(\objectiveFunctionSymbol | \dataset\right)$. We denote $\fmin$ as the smallest value observed of the objective $\objectiveFunction{\boldx}$ so far. Also, we use the substitution $z=(\performanceChangeSymbolMin-\performanceChangeSymbolDistributionMean{\boldx})/\performanceChangeSymbolDistributionStd{\boldx}$, where the $\performanceChangeSymbolDistributionMean{\boldx}$ and $\performanceChangeSymbolDistributionStd{\boldx}$ are the mean and standard deviation of the surrogate posterior predictive model $p\left(\objectiveFunctionSymbol | \dataset\right)$. Lastly, we denote $\Phi$ and $\phi$ as the cumulative distribution and probability density functions of a standard normal distribution $\mathcal{N}\left(0, 1\right)$. It then follows that:

\begin{equation} \label{difference_estimator_variance_estimator_proof}
\begin{split}
\ExpectedImprovement{\boldx} &= \EperformanceChange{\text{I}(\boldx)} \\
&= \EperformanceChange{\max\{\performanceChangeSymbolMin-\performanceChangeSymbolNoX, 0\}} \\
&= \EperformanceChange{\{\performanceChangeSymbolMin-\performanceChangeSymbolNoX\}\Indicator{\performanceChangeSymbolNoX<\performanceChangeSymbolMin}} \\
&= \int_{-\infty}^{\infty}\{\performanceChangeSymbolMin-\performanceChangeSymbolNoX\}\Indicator{\performanceChangeSymbolNoX<\performanceChangeSymbolMin}\performanceChangeSymbolDistribution \performanceChangeSymbolDifferential \\
&= \int_{-\infty}^{\performanceChangeSymbolMin}\{\performanceChangeSymbolMin-\performanceChangeSymbolNoX\}\,\performanceChangeSymbolDistribution\performanceChangeSymbolDifferential \\
&= \performanceChangeSymbolMin\int_{-\infty}^{\performanceChangeSymbolMin}\performanceChangeSymbolDistribution\performanceChangeSymbolDifferential -\int_{-\infty}^{\performanceChangeSymbolMin
}\performanceChangeSymbolNoX\; \performanceChangeSymbolDistribution \performanceChangeSymbolDifferential \\
&=\performanceChangeSymbolMin\Phi\left(\frac{\performanceChangeSymbolMin-\performanceChangeSymbolDistributionMean{\boldx}}{\performanceChangeSymbolDistributionStd{\boldx}}\right) -\performanceChangeSymbolDistributionMean{\boldx}\Phi\left(\frac{\performanceChangeSymbolMin-\performanceChangeSymbolDistributionMean{\boldx}}{\performanceChangeSymbolDistributionStd{\boldx}}\right) + \performanceChangeSymbolDistributionStd{\boldx}\phi\left(\frac{\performanceChangeSymbolMin-\performanceChangeSymbolDistributionMean{\boldx}}{\performanceChangeSymbolDistributionStd{\boldx}}\right) \\
&=\left[\performanceChangeSymbolMin-\performanceChangeSymbolDistributionMean{\boldx}\right]\Phi\left(\frac{\performanceChangeSymbolMin-\performanceChangeSymbolDistributionMean{\boldx}}{\performanceChangeSymbolDistributionStd{\boldx}}\right) + \performanceChangeSymbolDistributionStd{\boldx}\phi\left(\frac{\performanceChangeSymbolMin-\performanceChangeSymbolDistributionMean{\boldx}}{\performanceChangeSymbolDistributionStd{\boldx}}\right) \\
&= \performanceChangeSymbolDistributionStd{\boldx}\left[z\Phi\left(z\right) + \phi\left(z\right)\right]. \;\;\qed
\end{split}
\end{equation}
 Next, we derive the variance of the improvement function needed for the scaled expected improvement (SEI) acquisition function. We make use of the properties $\standardDistributionFirstDer =-\standardSymbol\standardDistributionShort$ and $\standardDistributionSecDer = (\standardSymbol^2-1)\standardDistributionShort$ of the standard Gaussian pdf $\standardDistributionShort$. We use same denotation as with EI:

\begin{equation}
    \begin{split}
        \VperformanceChange{\text{I}(\boldx)} &= \EperformanceChange{\text{I}^2(\boldx)}-(\EperformanceChange{\text{I}(\boldx)})^2 \\
        &= \EperformanceChange{\max\{\performanceChangeSymbolMin-\performanceChangeSymbol{\boldx},0\}^2} - (\text{EI}(\boldx))^2 \\
        &= \int_{-\infty}^{\performanceChangeSymbolMin} \{\performanceChangeSymbolMin-\performanceChangeSymbolNoX\}^2\performanceChangeSymbolDistribution\performanceChangeSymbolDifferential - (\text{EI}(\boldx))^2 \\
        &= \int_{-\infty}^{\standardSymbolMin} \{\performanceChangeSymbolMin - \performanceChangeSymbolDistributionMean{\boldx}-\performanceChangeSymbolDistributionStd{\boldx}\standardSymbol\}^2\standardDistributionShort\standardDifferential - (\text{EI}(\boldx))^2 \\
        &= \int_{-\infty}^{\standardSymbolMin} \{\left[\performanceChangeSymbolMin - \performanceChangeSymbolDistributionMean{\boldx}\right]^2+\performanceChangeSymbolDistributionVar{\boldx}\standardSymbol^2 -2\performanceChangeSymbolDistributionStd{\boldx}\standardSymbol\left[\performanceChangeSymbolMin - \performanceChangeSymbolDistributionMean{\boldx}\right]\}\standardDistributionShort\standardDifferential - (\text{EI}(\boldx))^2 \\
        &= \left[\performanceChangeSymbolMin - \performanceChangeSymbolDistributionMean{\boldx}\right]^2 \int_{-\infty}^{\standardSymbolMin} \standardDistributionShort\standardDifferential + \performanceChangeSymbolDistributionVar{\boldx}\int_{-\infty}^{\standardSymbolMin} \standardSymbol^2 \standardDistributionShort\standardDifferential \\ & -2\performanceChangeSymbolDistributionStd{\boldx}\left[\performanceChangeSymbolMin - \performanceChangeSymbolDistributionMean{\boldx}\right]\int_{-\infty}^{\standardSymbolMin}\standardSymbol\,\standardDistributionShort\standardDifferential - (\text{EI}(\boldx))^2 \\ &= \left[\performanceChangeSymbolMin - \performanceChangeSymbolDistributionMean{\boldx}\right]^2\Phi(\standardSymbolMin) + \performanceChangeSymbolDistributionVar{\boldx}\left[\int_{-\infty}^{\standardSymbolMin} \left(\standardSymbol^2-1\right) \standardDistributionShort\standardDifferential + \int_{-\infty}^{\standardSymbolMin} \standardDistributionShort\standardDifferential\right] \\ 
        & +2\performanceChangeSymbolDistributionStd{\boldx}\left[\performanceChangeSymbolMin - \performanceChangeSymbolDistributionMean{\boldx}\right]\int_{-\infty}^{\standardSymbolMin}-\standardSymbol\,\standardDistributionShort\standardDifferential - (\text{EI}(\boldx))^2 \\
        &=  \left[\performanceChangeSymbolMin - \performanceChangeSymbolDistributionMean{\boldx}\right]^2\Phi(\standardSymbolMin) + \performanceChangeSymbolDistributionVar{\boldx}\left[-\standardSymbolMin\standardDistributionWithSymbol{\standardSymbolMin} + \Phi(\standardSymbolMin)\right] \\ & +2\performanceChangeSymbolDistributionStd{\boldx}\left[\performanceChangeSymbolMin - \performanceChangeSymbolDistributionMean{\boldx}\right]\standardDistributionWithSymbol{\standardSymbolMin} - (\text{EI}(\boldx))^2 \\
&= \{ \left[\performanceChangeSymbolMin - \performanceChangeSymbolDistributionMean{\boldx}\right]^2 + \performanceChangeSymbolDistributionVar{\boldx}\}\Phi(\standardSymbolMin) - \performanceChangeSymbolDistributionStd{\boldx} \{\performanceChangeSymbolMin-\performanceChangeSymbolDistributionMean{\boldx}\}\standardDistributionWithSymbol{\standardSymbolMin}
\\ & +2\performanceChangeSymbolDistributionStd{\boldx}\left[\performanceChangeSymbolMin - \performanceChangeSymbolDistributionMean{\boldx}\right]\standardDistributionWithSymbol{\standardSymbolMin} - (\text{EI}(\boldx))^2 \\
&= \{ \left[\performanceChangeSymbolMin - \performanceChangeSymbolDistributionMean{\boldx}\right]^2 + \performanceChangeSymbolDistributionVar{\boldx}\}\Phi(\standardSymbolMin) + \performanceChangeSymbolDistributionStd{\boldx} \{\performanceChangeSymbolMin-\performanceChangeSymbolDistributionMean{\boldx}\}\standardDistributionWithSymbol{\standardSymbolMin} - (\text{EI}(\boldx))^2 \\
&= \performanceChangeSymbolDistributionVar{\boldx}\{(\standardSymbolMin^2+1)\Phi(\standardSymbolMin)+\standardSymbolMin\standardDistributionWithSymbol{\standardSymbolMin}\} - (\text{EI}(\boldx))^2.\;\;\qed
    \end{split}
\end{equation}

\end{document}